\documentclass{jfm}
    \usepackage[ pdftex, plainpages = false, pdfpagelabels, 
                 pdfpagelayout = useoutlines,
                 bookmarks,
                 bookmarksopen = true,
                 bookmarksnumbered = true,
                 breaklinks = true,
                 linktocpage,
                 pagebackref,
                 colorlinks = false,  
                 linkcolor = blue,
                 urlcolor  = blue,
                 citecolor = red,
                 anchorcolor = green,
                 hyperindex = true,
                 hyperfigures
                 ]{hyperref}

\shortauthor{T. Liu, B. Semin, L. Klotz, R. Godoy-Diana, J. E. Wesfreid and T. Mullin}

\title{Decay of streaks  and  rolls in plane Couette-Poiseuille flow}

\author{T. Liu\aff{1}
  , B. Semin\aff{1}\corresp{\email{benoit.semin@espci.fr}},
  L. Klotz\aff{2},
  R. Godoy-Diana\aff{1},
  J. E. Wesfreid\aff{1}, 
  \and T. Mullin\aff{3} 
 }

\affiliation{\aff{1}PMMH, CNRS, ESPCI Paris, Universit{\'e} PSL, Sorbonne Univ., Univ. de Paris, F-75005, Paris, France
\aff{2}Institute of Science and Technology, Am Campus 1, 3400 Klosterneuburg, Austria
\aff{3}Mathematical Institute, University of Oxford, Oxford, OX2 6GG, UK}

\begin{document}
%
%
%
%
%
%

\maketitle

\begin{abstract}
	
We report the results of an experimental investigation into the decay of turbulence in plane Couette-Poiseuille flow using `quench' experiments where the flow laminarises after a sudden reduction in Reynolds number $\Rey$. 
Specifically, we study the velocity field in the streamwise-spanwise plane.
We show that the spanwise velocity containing  
rolls, decays faster than the streamwise velocity, 
 which displays elongated regions of higher or lower velocity called streaks. At final Reynolds numbers above $425$, the decay of streaks displays two stages: first a slow decay when rolls are present and secondly a more rapid decay of streaks alone. The difference in behaviour results from the regeneration of streaks by rolls, called the lift-up effect. We define the  turbulent fraction as the portion of the flow containing turbulence and this is estimated by thresholding the spanwise velocity component. It  decreases linearly with time in the whole range of final $\Rey$. The corresponding decay slope increases linearly with final $\Rey$. The extrapolated value at which this decay slope vanishes is $\Rey_{a_z}\approx 656\pm10$, close to $\Rey_g\approx 670$ at which turbulence is self-sustained. The decay of the energy computed from the spanwise velocity component is found to be exponential. The corresponding decay rate increases linearly with $\Rey$, with an extrapolated vanishing value at $\Rey_{A_z}\approx 688\pm10$. This value is also close to the value at which the turbulence is self-sustained, showing that valuable information on the transition can be obtained over a wide range of $\Rey$. 

\end{abstract}

%

\section{Introduction}

The transition to turbulence is complex in wall-bounded shear flows. Examples include plane Couette flow (PCF), plane Poiseuille flow (PPF) and Couette-Poiseuille flow (CPF). The transition scenario in these flows is usually termed subcritical and characterized by the coexistence of turbulent and laminar regions in the transition regime. In PCF and CPF experiments where the flow is driven by a moving belt, finite amplitude background disturbances are inevitably present and we will refer to them as 'noise' in this article. Even if the noise is small in amplitude, the transition to turbulence occurs at values of $\Rey$ that are finite and thus lower than the theoretical linear critical Reynolds number $\Rey_l$, which is infinite for PCF and CPF with zero mean flow \citep{klotz_wesfreid_2017jfm}. For plane shear flows induced by pressure gradients such as PPF,  careful design of the setup can give transition around $\Rey_l=5772$ \citep{Orszag1971} (see the definition of $\Rey$ below).

Our focus is on the transition to turbulence in plane Couette-Poiseuille flow, where the flow is driven by one sided shear and the mean flux is approximately zero. It is the simplest Couette-Poiseuille flow to realize experimentally \citep{tsanis_leutheusser_1988}. It can be considered as  intermediate between the widely studied cases of plane Couette and plane Poiseuille flows: the main component of the motion is Couette-shear with a weak  return  Poiseuille flow (see figure~\ref{fig:setup}). We will investigate the transition process by 'quenching', i.e. sudden decrease in $\Rey$.

We first discuss the transition in PCF and PPF. Plane Couette flow has been extensively studied experimentally and numerically. The Reynolds number is defined using the belt velocity $U_{belt}$ and the half-gap $h$: $\Rey= hU_{belt}/\nu$ where $\nu$ the kinematic viscosity of the fluid.
The global stability threshold is the Reynolds number above which the turbulent state is sustained. It is denoted by $\Rey_g$ in the present article (the notation $Re_c$ is also used in the literature). 
 The value $\Rey_g=323 \pm 2$ was determined experimentally by \cite{BottinChate1998}, and $\Rey_g=324 \pm 1$ was established numerically by \cite{Duguet2010JFM} in large domains. In order to avoid any arbitrariness in the choice of the lifetime for decay under quenching, the numerical work by \cite{LiangShi2013PRL_PCF} used an approach based on the equality of the splitting and decay rates of the turbulent regions. These are found to form banded structures in the decay regime. They found a similar global stability threshold $\Rey_g=325$ despite using a narrow tilted domain with respect to streamwise direction introduced by \cite{Barkley2005}. Another characteristic threshold is the Reynolds number $\Rey_t$ at which turbulence becomes featureless, which is larger than $\Rey_g$. It has been estimated experimentally as $\Rey_t=415$ by \cite{Prigent2001} and numerically as $\Rey_t=420$ by \cite{Duguet2010JFM}. The behaviour close to $\Rey_t$ was studied numerically in detail by \cite{Rolland_2018}, showing that several crossover Reynolds numbers can be defined close to $\Rey_t$. The transitions of laminar-turbulent bands to featureless turbulence occurs at  each of the crossover points.

The behaviour in plane Poiseuille flow is more complex than PCF. The Reynolds number is usually defined using the center-plane velocity of the corresponding laminar flow and the half-gap $h$. The global threshold obtained numerically is $\Rey_g =700$ \citep{Shimizu2019}. This is consistent with the experimental results of PPF  by \cite{Paranjape2019} where a positive mean growth rate of the turbulent bands for $\Rey>650$ is found in quench experiments \citep{BottinChate1998, Monchaux2020jfm}. Several crossover Reynolds numbers can be defined, linked to the existence of bands and of their orientation. A 'lower marginal Reynolds number'  $\Rey=1050$ is obtained from the linear extrapolation of the intermittency factor by \cite{Seki2012_pof}. By way of contrast with PCF, the crossover Reynolds number obtained by equating the decay and splitting rate in a tilted narrow channel ($\Rey=965$, see the numerical work by \cite{Sebastien2020PRF}) is different from $\Rey_g$.

Since the mean flux is approximately zero in our experiment we use the belt speed $U_{belt}$ as the characteristic velocity. The Reynolds number can thus be defined using this and the half-gap $h$: $\Rey= hU_{belt}/\nu$. This configuration has not been explored in as much detail as either PCF or PPF. The threshold at which turbulence is self-sustained in experiments is estimated approximately at $\Rey\approx 670$; the turbulence becomes featureless $\Rey\approx780$ \citep{Klotz2017PRF}. Since the geometry of CPF is similar to that of PCF and PPF, it is anticipated that there will be common features in the transition processes in all three flows. 

An important geometric parameter is the aspect ratio, i.e. the size of the channel in the streamwise and spanwise directions relative to the half channel width $h$. Turbulence cannot be sustained below a size called the `minimal flow unit' \citep{jimenez_moin_1991,hamilton_kim_waleffe_1995}, where the width and height are a few $h$ in wall-bounded flow. The influence of the aspect ratio has been investigated numerically in PCF by \cite{Manneville2011PRE_SizeEffect_DecaySnapshots,Rolland_2018}. They show that the channels of sizes below around $80h$ only display temporal dynamics, while above $80h$ both spatio and temporal dynamics can be captured. For such channels, turbulent bands aligned at a well-defined angle with the streamwise direction are separated by laminar regions. The wavelength of such bands is of order $70-80h$ \citep{Manneville2011PRE_SizeEffect_DecaySnapshots}.
Characteristics of flows in infinite domains are obtained in practice only for very large sizes ($2000h$), which have been studied using models with truncated equations \citep{chantry_tuckerman_barkley_2017}. In this case, the turbulent fraction, defined as the ratio of the turbulent region with respect to the entire area, is a continuous function of $\Rey$ in the turbulent steady state. The minimal size of the domain required to observe complex spatio-temporal behaviours in CPF is not yet  known. The aspect ratio of our experiment is sufficiently large to observe oblique bands.

As noted above, one way to study the properties of turbulence is to investigate its decay using quench experiments, i.e. the transition from turbulent to laminar flow \citep{Batchelor1948, BottinChate1998,Prigent2005,PeixinhoetMullin2006PRL,Rolland2015, Paranjape2019}. The advantage of the quench protocol is that the flow is initialized in the fully turbulent state which is less sensitive to external noise than the laminar state. Changing the flow rate rapidly is challenging experimentally in Poiseuille flows, but it is relatively straightforward in our experiment where the flow is driven by a belt. 

We have carried out an experimental investigation of the decay of the streamwise and spanwise components of the velocity field and highlight their roles in the relaminarization process. In addition, these two components provide information concerning the structures that drive the self-sustained cycle of turbulence \citep{Waleffe1997SSP}: on the one hand, the modulation of the streamwise velocity gives rise to the structures called \emph{streaks}, and on the other hand the spanwise velocity characterises the dynamics of streamwise vortices--also called \emph{rolls}-- which accompany the streak  dynamics.

The difference in behaviour of these two components has been discussed by several authors: in a simplified model used by \cite{Rolland2018PRE}, the proxies for the streamwise and spanwise components display different behaviours. The different decay rates of the velocity components during turbulent decay described in this work also received attention in the recent numerical work by \cite{Sebastien2020PRF}.
The different behaviors associated with the various flow components have also been investigated in the permanent regime by \cite{Duriez2009} for a flat-plate boundary layer.

In the current investigation, the decay of the streamwise and spanwise velocities is carried out over a large range of Reynolds numbers, ranging from approximately half to slightly larger than $\Rey_g$. This is in contrast with many previous studies, which focus on values of $\Rey$ very close to or slightly above $\Rey_g$. We aim at showing that the decay rate difference is observed over a wide range of $\Rey$, and that the global threshold $\Rey_g$ is close to the value extrapolated from the value of the decay rate at small $\Rey$. 
Another objective of the present study is to investigate the interplay between the rolls and streaks in the flow, in particular to elucidate the dynamics of each component.

In our Couette-Poiseuille setup, noise is generated in the fluid supply tank and disturbed flow is thus injected into one end of the channel (see figure~\ref{fig:setup}). Similar behaviour is observed in experiments on torsional Couette flow \citep{LeGal2007JEM_CircularPCF} and PPF \citep{sano2016}.  Disordered flow may penetrate the flow field from both end tanks in CPF, since our experiment has a supply tank at each end \citep{Couliou2015PhysFluids}.
Contrary to the case of boundary layer flows, which is another example of highly sheared flow where the effect of noise has been investigated
\citep{fransson2005, Kreilos2016}, it is not common to vary the noise in channel flows.
Here we characterised the noise and controlled it using grids. 

The article is organized as follows. We present the main features of the experimental setup, the velocity measurements and the processing steps in section~\ref{Experimental set-up and processing}. The spatial structure of the velocity fields during the relaminarization process, as well as the temporal evolution of characteristic integral parameters such as kinetic energy and turbulent fraction are discussed in section~\ref{Decay process}. The noise is characterised and quantified in section~\ref{Noise}, using the velocity field in the permanent regime. The variation with Reynolds number of the characteristic decay times is discussed in section~\ref{Variation of characteristic time with $Re$}.

\section{Experimental set-up and processing}
\label{Experimental set-up and processing}

\subsection{Experimental set-up}

\begin{figure}
  \centerline{\includegraphics[width=13cm]{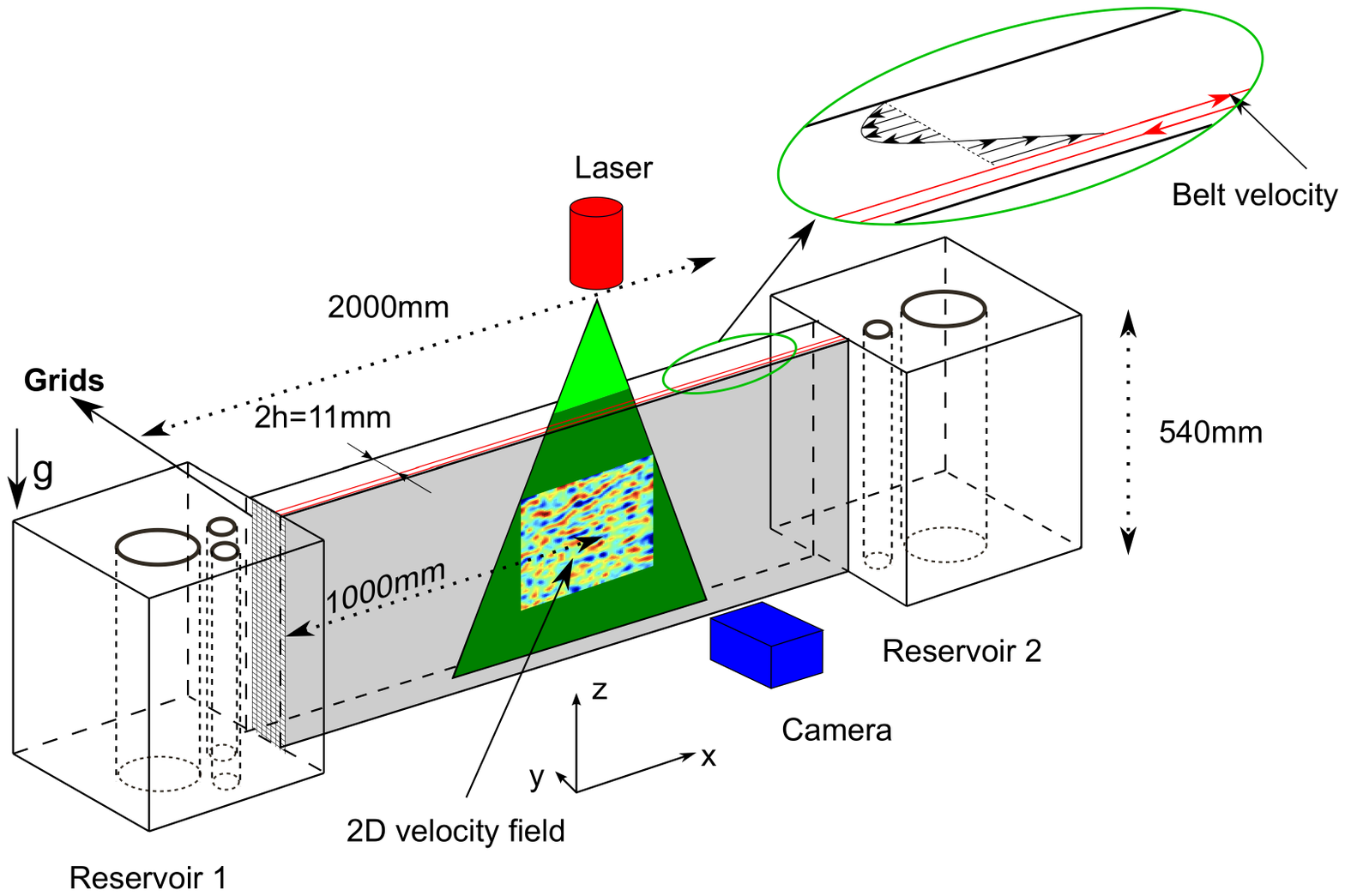}}
  \caption{A schematic diagram of the experiment.}
\label{fig:setup}
\end{figure}
A schematic diagram of the apparatus is shown in figure~\ref{fig:setup}. It has previously been described in detail by \cite{Klotz2017PRF}. It consisted of two parallel vertical glass plates set $14$ mm apart and these form a connected channel between two water reservoirs filled. The glass plates were closed at the top and at the bottom by two horizontal surfaces, forming a channel. The tops of the reservoirs were not closed. The set-up was filled with water at room temperature $21.5^\circ$C$\pm1.5^\circ$C and the viscosity of the water was evaluated at the measured water temperature. 

The belt was a Mylar membrane which was guided by vertical cylinders so that it was parallel to the vertical glass plates, and close to one of the plates. One of the cylinders in reservoir~1 rotates, which produces a translation motion of the membrane at constant velocity $U_{belt}$.

The flow of interest was in the widest gap between the moving membrane and the fixed glass plate. For consistency with previous investigations, the width of this gap is defined as $2h$, where $2h=11.0 \pm 0.3$~mm. The belt velocity $U_{belt}$ created a shear flow which also induced a pressure difference between the two reservoirs. This pressure difference created a counter flow, so that the mean flow was almost zero in the wall-normal $y$ direction. A parabolic (Couette-Poiseuille) profile was obtained in the laminar regime.

The length of the channel in the streamwise direction, i.e. the $x$ direction, is $L_x=2000$ mm, so that $L_x/h=364$. The height of the channel in the spanwise direction,  i.e. the $z$ direction, is $L_z=540$ mm, so that $L_z/h=98$.
The half channel width $h$ and the belt velocity $U_{belt}$ are used to make the variables dimensionless. The Reynolds number is defined as $\Rey=U_{belt}h/\nu$, where $\nu$ is the kinematic viscosity of water with $\nu~\in~[0.934, 1.003]$~ mm\textsuperscript{2}$\cdot$s\textsuperscript{-1} in our experiments. The $U_{belt}$ was in the range of $[0.03, 0.2]$ m$\cdot$s\textsuperscript{-1} in our study. In the following of the article, we will denote dimensional parameters by an asterisk exponent.

The rotating cylinder in reservoir $1$ which drove the belt induced a large-Reynolds-number turbulent flow in the reservoir. At $\Rey=600$ in the channel, the $\Rey$ in reservoir $1$ was approximately $2\times10^4$ which is calculated using the half width of the reservoir $H^*=23$ cm and $U_{belt}$. This source of turbulence acted as external noise for the flow inside the channel. Some of the turbulence generated in the reservoirs invaded the flow channel as reported in PCF experiments \citep{BottinChate1998, Couliou2015PhysFluids}.

A novelty of the present experiment was the addition of multi-layer grids at the junction between reservoir~1 and the channel to help reduce the noise that perturbs the flow in the channel. Fine mesh grids have previously been used in boundary layer flows to reduce the streaky flow and homogenise the incoming flow \citep{Puckert2017BLgrids}. 
The multi-layer grids consisted of $5$ stainless steel grids with a distance between the layers of $1-2$~mm. The diameter of the wires was $0.4$ mm. The size of the grids was $25$ (width) $\times$ $500$ (height) mm. The mesh size $1$ mm was significantly smaller than $2h$ and breaks up the large eddies and prevents them from entering the channel. It was found that the level of noise in the channel was sensitive to the exact position of the grid. We studied four levels of external noise: one without the grid (high noise) and three with the grid in place.

\subsection{Particle image velocimetry}
\label{PIV}
Two-dimensional Particle Image Velocimetry (PIV) was used to measure the velocity field in the $xz$ plane. The location of this plane in the $y$ direction was $y=0.33\pm0.04$, which is the position where the velocity passes through zero in the laminar profile (see figure~5($b$) \citep{Klotz2017PRF}). This plane was selected using a laser sheet obtained from a Darwin-Duo\textsuperscript{\textregistered} 20 mJ Nd-YLF double-pulse green laser (527 nm). The time interval between the two laser pulses was $\Delta t^* = 12.5$ ms and the pulse duration was less than $250$ ns. The fluid was seeded to enable PIV with particles of diameter $20$ $\mu$m made of polyamide (density $1.03$ g$\cdot$cm$^{3}$) with a volume concentration of $1.7\times 10^{-5}$ g$\cdot$ml$^{-1}$ . 

Images were acquired using a camera Imager MX5M\textsuperscript{\textregistered} from LaVision\textsuperscript{\textregistered} ($2464 \times 2056$ pixels) with a frequency $f^*=2$ Hz using the double frame mode. The time duration between two consecutive frames was set as the interval between two laser pulses. A Nikon\textsuperscript{\textregistered} objective lens $17-35$ mm with an aperture $f/2.8$ was mounted with a distance $920$ mm from the measurement plane. 
The field of view was fixed at the middle of the channel, around $180h$ between the center of the measurement field and the entrance of the channel from the reservoir $1$ side (see figure~\ref{fig:setup}). The size of the measurement field was $77h \times 79h$.

The velocity fields were computed using DaVis 10 software (LaVision) with a multi-pass algorithm. As the velocity field was dominated by the streamwise velocity component, the displacement of the particles in this direction was more than one order of magnitude larger than the spanwise. Therefore, we used an interrogation window which is elliptical with an aspect ratio  $4:1$ between the streamwise and spanwise directions. The total number of pixels of this interrogation windows is $2304$, and the overlap between two successive windows is $50\%$. As the PIV calculation induces some artifacts close to the boundary of the measured field, the velocity field was cropped to a size of $65h\times 67h$.

\subsection{Protocol}

The following protocol was used in the experiments: the flow was initialized at $\Rey_i = 1000 > Re_t$, i.e. in the featureless turbulent regime. The belt speed was then suddenly reduced to the lower final Reynolds number $\Rey_f$. This protocol is commonly referred to as a quench experiment \citep{BottinChate1998, Monchaux2020jfm}. The decrease of the Reynolds number was achieved by decreasing the velocity of the membrane, using a Labview program controlling the rotation of the motor as a function of time. The time required to change the belt velocity is less than $0.1$~s, i.e. at most $2$ time units ($h/U_{belt}$). This time is much smaller than the typical decay time of the turbulence in the channel. In the following, time $t=0$ corresponds to the time at which the Reynolds number is decreased.

\subsection{Small scales}

The velocity $U$ can be decomposed into $U = u_{lsf}+u$, where $u_{lsf}$ is the large-scale flow (LSF) and $u$ is the small-scale flow (SSF). Large scale flows in wall-bounded shear flows arise from the non-zero spanwise velocity component \citep{Duguet2013}, and a small contribution from the imperfections of the membrane in the channel. The scale separation in the present set-up was investigated by \cite{klotz_wesfreid_2020jfm}. 

In this investigation, we remove the LSF and focus on the SSF $u$, which is the most significant contribution to the turbulent flow field \citep{lemoult_2013}. We used a $2D$ fourth-order Butterworth spatial filter with a cutoff wavelength $\lambda\leq 14.8$ to remove large scale flows. The wavelength $\lambda$ is defined as $2\upi/\lambda=k=\sqrt{k_x^{2}+k_z^{2}}$, where $k_x$ and $k_z$ are the streamwise and spanwise wavenumber, respectively. The results do not change qualitatively when $\lambda$ is varied between $8.4$ and $16.8$. For example, the decay time $\tau$ (defined in Section 5) changed by less than $5\%$ for measurements at final Reynolds number $\Rey_f=500$.  The small-scale velocity fluctuation $u_x$  is a measure of the streaks and the spanwise velocity $u_z$ corresponds to the streamwise vortices, also termed rolls.

\subsection{Energy and turbulent fraction}

We characterise the global state of the flow in the field of view using both kinetic energies and turbulent fractions. To investigate possible different behaviours of the velocity in the streamwise  ($u_x$) and spanwise ($u_z$) directions, we define variables which only depend on either of these velocity components. We recall that $u_x$ is one order of magnitude larger than $u_z$. This approach was used by \cite{klotz_wesfreid_2017jfm} for the study of transient growth in CPF.

We define the streamwise `perturbation energy' $E_{x}$ as:
\begin{equation}
E_{x} = \frac{1}{2L_xL_z}\int_{-L_z/2}^{L_z/2} \int_{0}^{L_x}{u_x}^2dxdz
\label{Eq:ExDefinition}
\end{equation}
Similarly, we define the spanwise energy of the rolls $E_{z}$ as:
\begin{equation}
E_{z} = \frac{1}{2L_xL_z}\int_{-L_z/2}^{L_z/2} \int_{0}^{L_x}{u_z}^2dxdz
\label{Eq:ExDefinition2}
\end{equation}

\noindent Since we use a non-dimensional quantity, the density of the fluid is not explicitly involved in the definition of the energy.

Several methods have been used to estimate the turbulent fraction of the flow field defined as the fraction of space where the flow is turbulent. Experimentally or numerically, the velocity is often non-zero even in the laminar regions.  Hence, there is some arbitrariness in the choice of the variable which is used to define the turbulent fraction, as well as in the choice of the threshold.

Pioneering experiments on PCF used visualization of the flow with anisotropic iriodin particles (see \cite{tillmark_alfredsson_1992}, \cite{Daviaud1992PRL_PCF} and \cite{BottinChate1998}). This is an indirect measurement of the local velocity field. The energy averaged on a cell size close to that of the minimal flow unit has been used in the numerical work of \cite{RollandMannevilleEPJB2011}. The streamwise velocity is the dominant contribution to the energy, and thus to the latter definition of the turbulent fraction. In their experimental work, \cite{Monchaux2020jfm} use a method based on the measurement of the normal vorticity.
Despite these differences, the qualitative variation of the turbulent fraction  with time or Reynolds number is consistent.

Since we focus on the flow structures of the turbulent flow, we chose to define two 'turbulent fractions'. The turbulent fraction $F_x$ is computed from the streamwise velocity: a point is considered as turbulent if $|u_x| >1.4 \times 10^{-2}$. This value was obtained by comparing the velocity field and the turbulent region after thresholding. The typical $|u_x|$ of the streaks is around $6\times10^{-2}$.  Similarly, we define $F_z$ using the spanwise velocity only: a point is considered as turbulent if
$|u_z| >7 \times 10^{-3}$. The typical $|u_z|$ of the rolls is around $3\times10^{-2}$.

\section{Decay process}
\label{Decay process}


We outline typical features of the decay processes found in quench experiments using the results from two representative cases: one at $\Rey_f=425$, which is far below $Re_g$, and a second at $Re_f=600$, which is closer to this threshold. We also investigate the influence of the final Reynolds number on the decay process. 

\begin{figure}
  \centerline{\includegraphics[width=\textwidth]{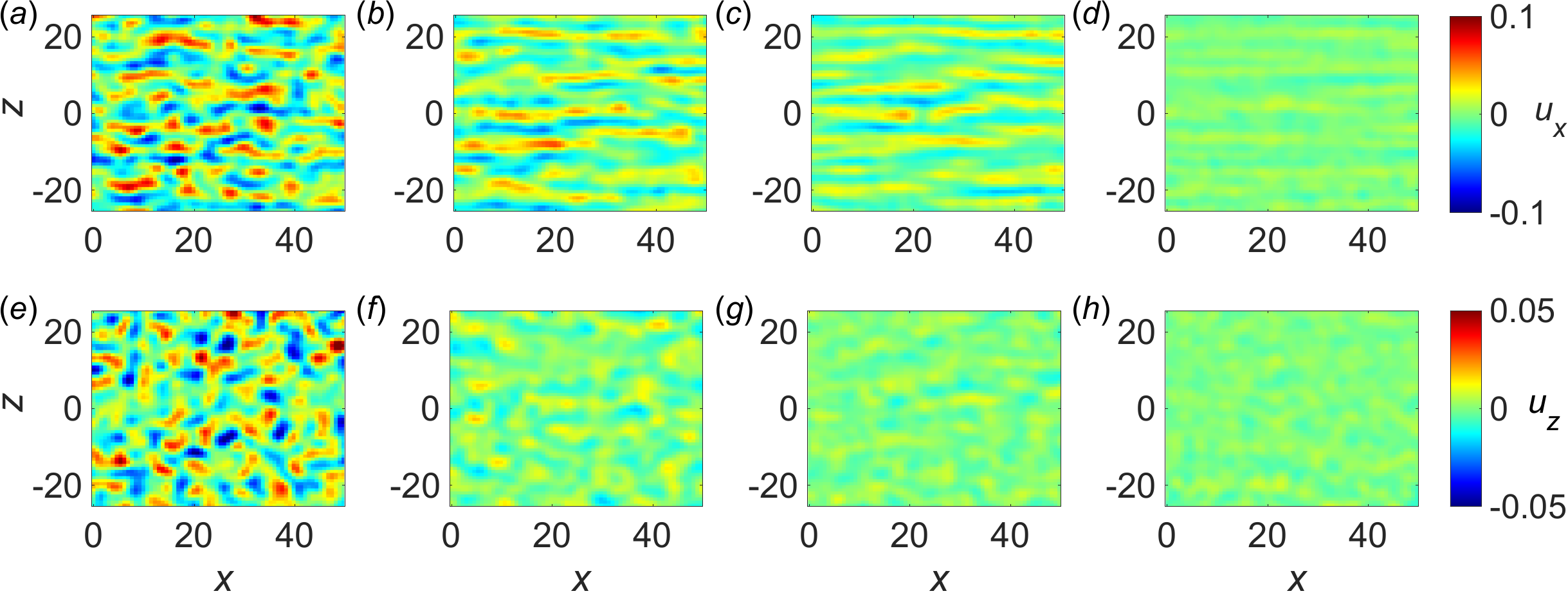}}
  \caption{Snapshots of velocity fields for different times at $Re_f=425$. Top row: velocity fields  in the streamwise direction $u_x$, bottom row: velocity fields in the spanwise direction $u_z$. Times: ($a$) and ($e$) $t=-65$ (fully turbulent flow, $\Rey_i=1000$), ($b$) and ($f$) $t=91$, ($c$) and ($g$) $t=150$, ($d$) and ($h$) $t=286$; noise intensity: high ($\sigma=4.6\times10^{8}$, defined in Section ~\ref{External noise in the permanent regime}).}
\label{fig:SnapRef425}
\end{figure}

Velocity fields for different times are shown in figure~\ref{fig:SnapRef425} for a $Re_f=425$ experiment: the top row ($a$-$d$) and bottom row ($e$-$h$) show, respectively, the streamwise $u_x$ and spanwise $u_z$ fields.  Figure~\ref{fig:SnapRef425} ($a$) and ($e$) are respectively the streamwise and spanwise velocity fields before the quenching, i.e. when the Reynolds number is $Re_i=1000$. As expected, the flow is fully turbulent. The streaks can be identified as the elongated structures aligned in the $x$ direction in figure~\ref{fig:SnapRef425} (a). These streaks have a typical length of $10h-20h$, and are typically not straight as in this figure. The velocity field $u_z$  displayed in figure~\ref{fig:SnapRef425} (e) is irregular, as expected for a turbulent flow. The magnitude of $u_z$ is one order of magnitude smaller than for $u_x$, which is a common feature of 3D flow structures in wall-bounded shear flows.

A typical evolution of the decay of turbulence at three successive time instants is displayed in figure~\ref{fig:SnapRef425}($b$-$d$) and ($f$-$h$). The velocity fields of $u_x$ after the quench are shown in figure~2($b$), ($c$) and ($d$). The streaks become longer and broader. The corresponding  $u_z$ velocity fields are shown in figure~\ref{fig:SnapRef425}($f$), ($g$) and ($h$). The decay of $u_z$ is faster than $u_x$, as can be seen for example at $t=150$, by comparing figure~\ref{fig:SnapRef425}($c$) and ($g$) and the shape of the structures in the $u_z$ field does not change significantly. The decay of the velocity field of $u_x$ is different from that of $u_z$. This decay scenario of streaks is qualitatively similar to that found in numerical simulations of PCF \citep{Manneville2011PRE_SizeEffect_DecaySnapshots}. This was attributed by them to a viscous damping effect and is typical for decaying turbulence \citep{Batchelor1948}.

\begin{figure}
  \centerline{\includegraphics[width=\textwidth]{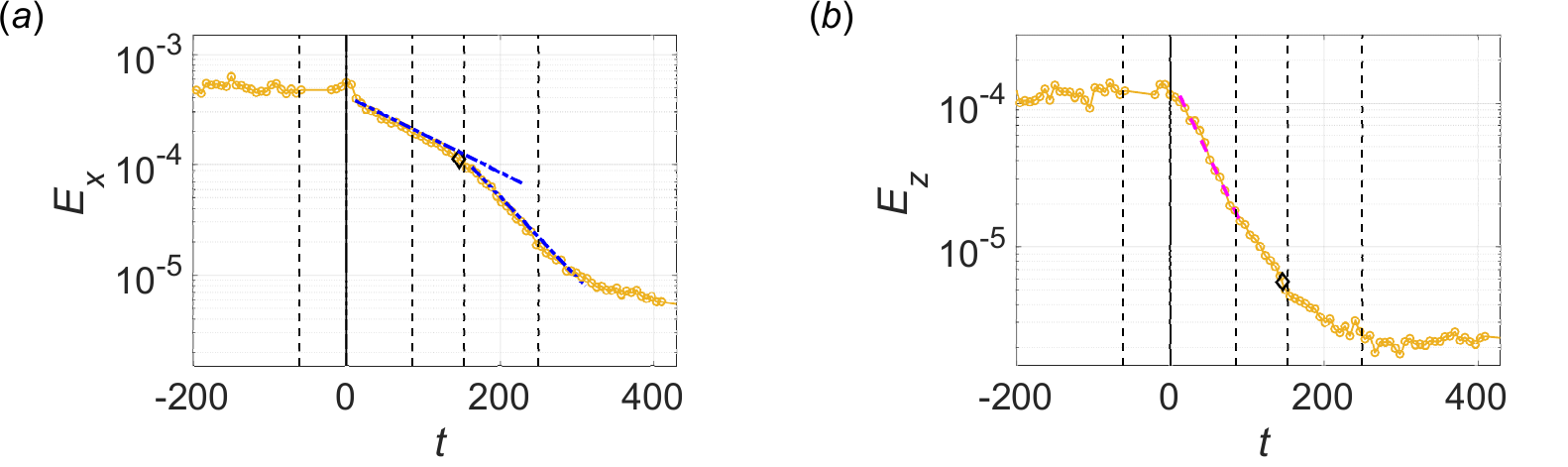}}
  \caption{Temporal evolution of $E_{x}$ in ($a$) and of $E_{z}$ in ($b$) for $\Rey_f=425$; dashed vertical lines represent the times for the snapshots of $u_x$ and $u_z$ plotted in figure~\ref{fig:SnapRef425}; blue dot-dashed lines: guide for the eyes to distinguish the different decay stages for $E_x$; magenta dashed line: exponential fits $E_{z}=E_0\exp(A_zt)$; black diamond: the position of $\tau_z$ when $E_z$ decreases to $5\%$ of its initial energy; noise intensity: high ($\sigma=4.6\times 10^{8}$, defined in Section ~\ref{External noise in the permanent regime}).}
\label{fig:ExEzRef425}
\end{figure}

\begin{figure}
  \centerline{\includegraphics[width=\textwidth]{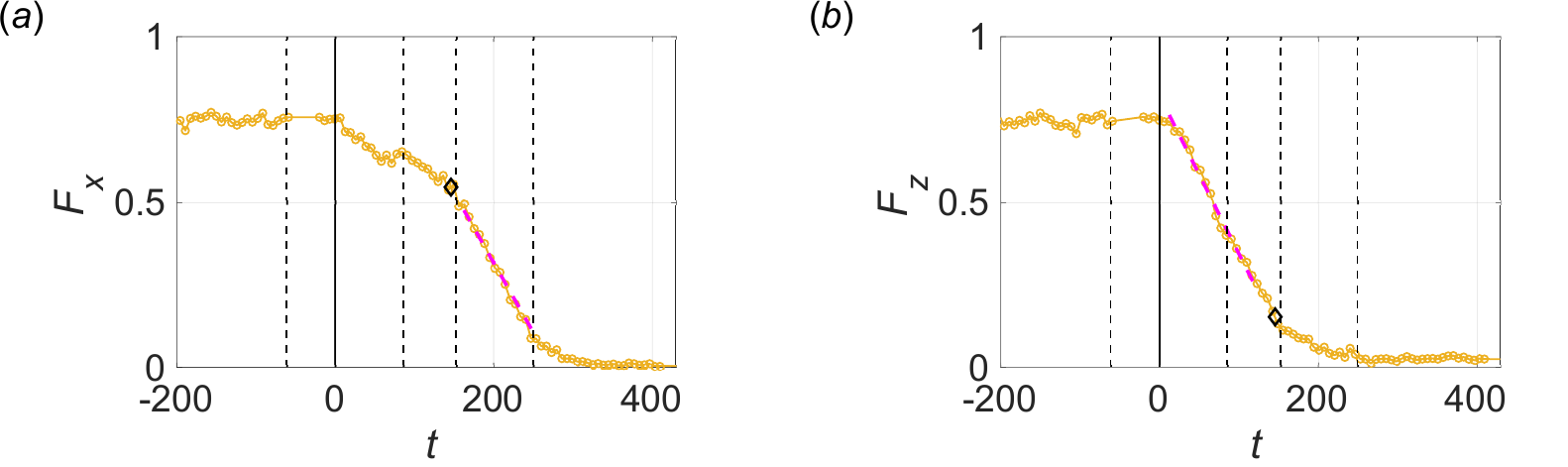}}
  \caption{Temporal evolution of $F_x$ in ($a$) and of $F_z$ in ($b$) for $\Rey_f=425$; dashed vertical lines represent the times for the snapshots of $u_x$ and $u_z$ plotted in figure~\ref{fig:SnapRef425}; magenta dashed lines lines: linear fits $F_{i}= a_{i}t+b$, ($i=x,z$); black diamond: the position of $\tau_z$ when $E_z$ decreases to $5\%$ of its initial energy; noise intensity: high ($\sigma=4.6\times10^{8}$, defined in Section ~\ref{External noise in the permanent regime}).}
\label{fig:FtxFtzRef425}
\end{figure}

\begin{figure}
  \centerline{\includegraphics[width=\textwidth]{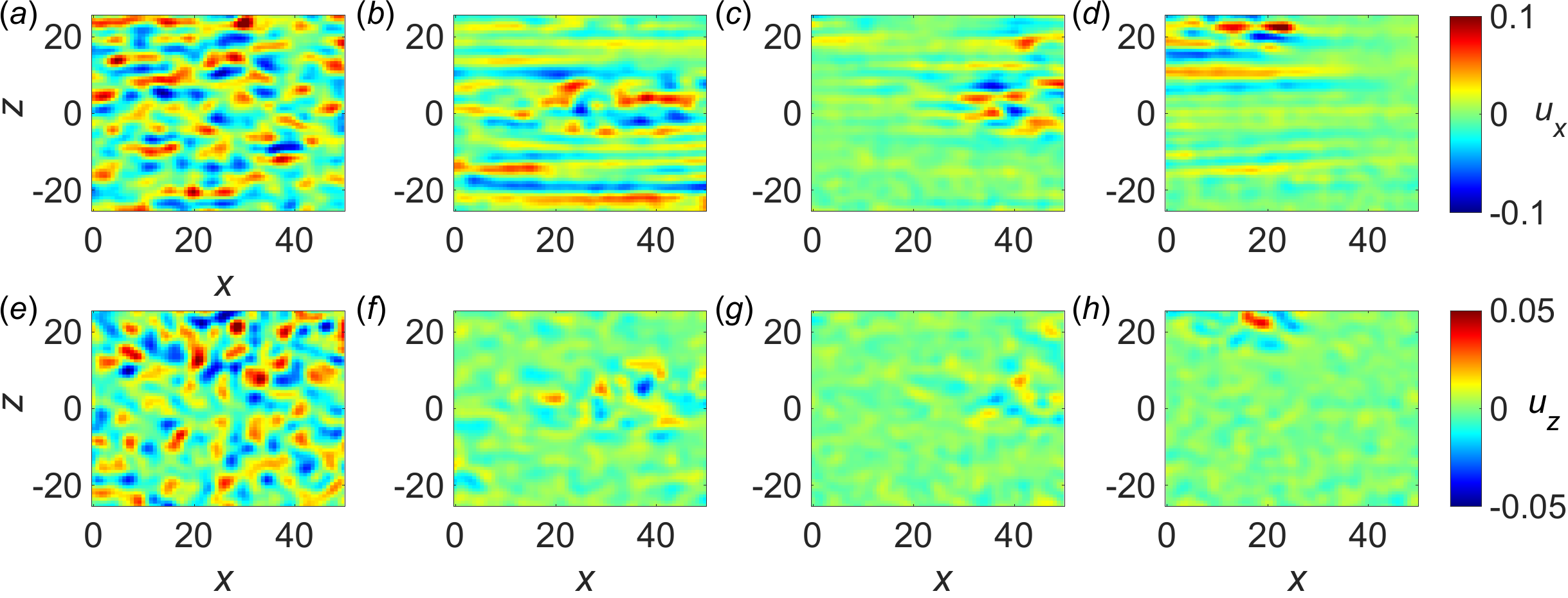}}
  \caption{Snapshots of $u_x$ in the top row and of $u_z$ in the bottom row at time for $\Rey_f=600$: ($a$) and ($e$) $t=-215$, ($b$) and ($f$) $t=430$, ($c$) and ($g$) $t=1255$, ($d$) and ($h$) $t=2430$; noise intensity: high ($\sigma=4.6\times10^{8}$, defined in Section ~\ref{External noise in the permanent regime}).}
\label{fig:SnapRef600}
\end{figure}

\begin{figure}
  \centerline{\includegraphics[width=\textwidth]{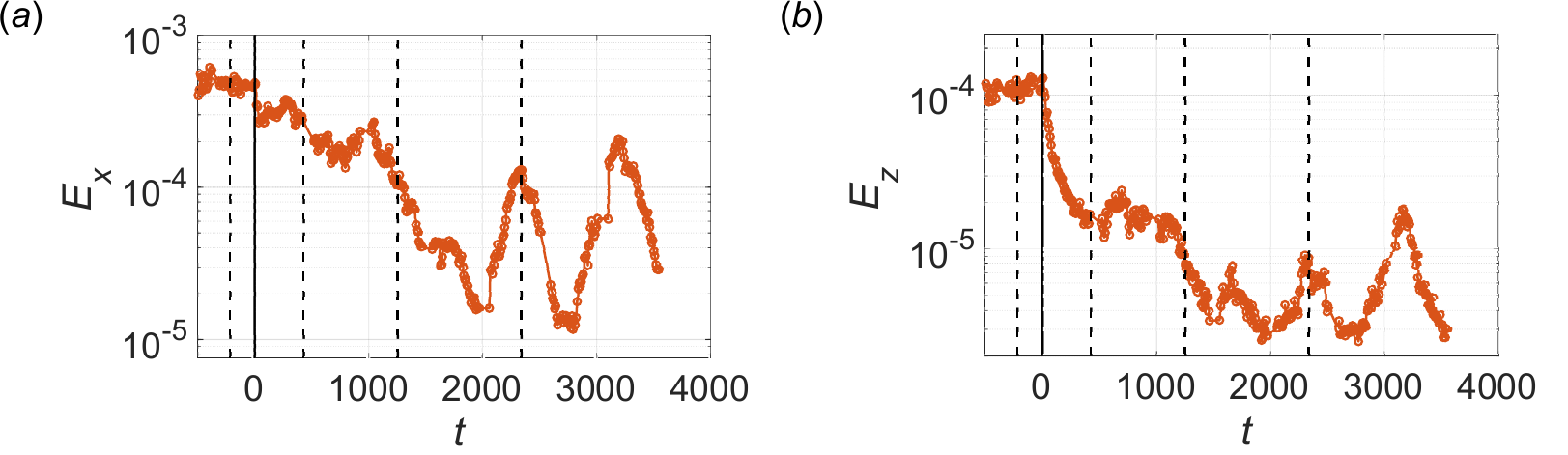}}
  \caption{Temporal evolution of $E_{x}$ in $(a)$ and of $E_{z}$ in ($b$) for $\Rey_f=600$; dashed vertical lines represent the times for the snapshots of $u_x$ and $u_z$ plotted in figure~\ref{fig:SnapRef425}; noise intensity: high ($\sigma=4.6\times10^{8}$, defined in Section ~\ref{External noise in the permanent regime}).}
\label{fig:ExEzRef600}
\end{figure}

\begin{figure}
  \centerline{\includegraphics[width=\textwidth]{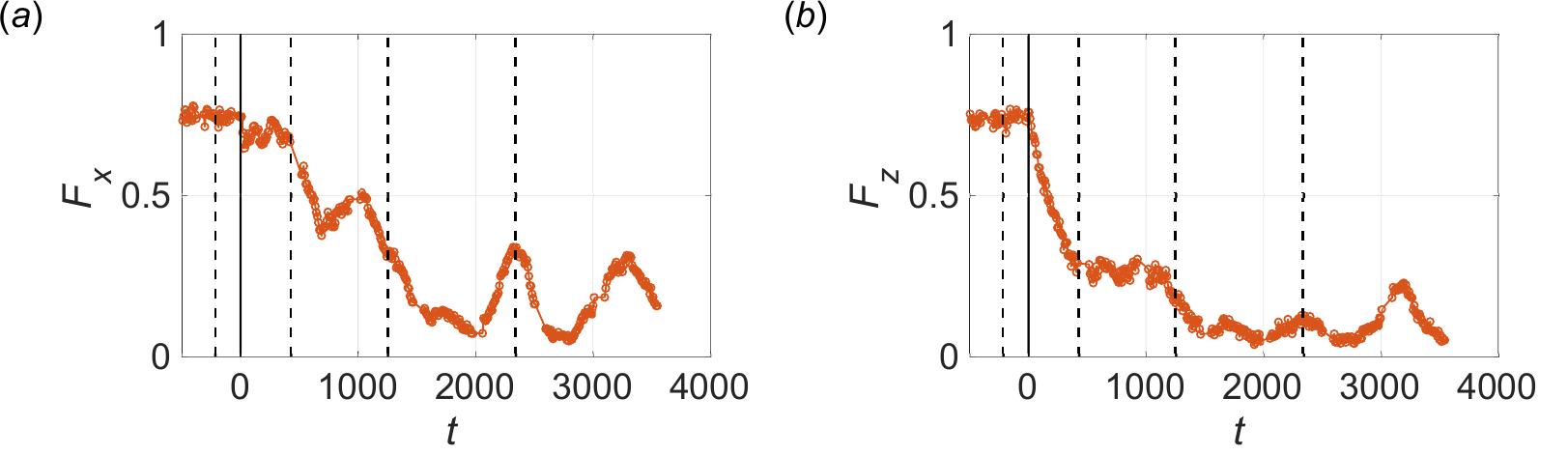}}
  \caption{Temporal evolution of $F_x$ in ($a$) and of $F_z$ in ($b$) for $\Rey_f=600$; dashed vertical lines represent the times for the snapshots of $u_x$ and $u_z$ plotted in figure~\ref{fig:SnapRef600}; noise intensity: high ($\sigma=4.6\times10^{8}$, defined in Section ~\ref{External noise in the permanent regime}).}
\label{fig:FtxFtzRef600}
\end{figure}

The temporal energy evolution for the streamwise component $E_{x}$ is shown in figure~\ref{fig:ExEzRef425}($a$) and the spanwise component $E_{z}$ in figure~\ref{fig:ExEzRef425}($b$). Note that the abscissa is linear, whereas the ordinate is displayed on a logarithmic scale. The dashed vertical black lines indicate the times at which the corresponding velocity fields are plotted in figure~\ref{fig:SnapRef425} to illustrate the dynamics of streaks and rolls. The magnitude of the energy depends on the wall-normal $y$ position of the measured field (see section~\ref{PIV}). The uncertainties of the position can change the absolute value of the measure but this does not have a significant influence on the results. In this example, the energies decrease monotonically, which is linked to the small value of the Reynolds number. $E_z$ decreases faster than the $x$-component, in agreement with the observation discussed above with reference to figure~\ref{fig:SnapRef425}. We observed two different decay stages in the evolution of $E_x$ after the quench:  ($1$) $t\lesssim 160$, the decay accompanied with elongating and flattening of streaks, which corresponded to the snapshots in figure \ref{fig:SnapRef425}($b$) and ($c$); ($2$) $t\gtrsim 160$, fading of streaks induced by viscous damping, which corresponded to the snapshots from figure \ref{fig:SnapRef425}($c$) to ($d$).
 
We also compared the decays of  $E_x$ and $E_z$ and found that $E_z$ is negligible during the second stage of the decay of $E_x$. To quantify this, we define the decay time $\tau_z$ at which the energy $E_z$ decreases to $5\%$ of its initial value $E_i$. The choice of the threshold for the definition of $\tau_z$ will be discussed in Section \ref{Variation of characteristic time with $Re$}. The data point at $\tau_z$ is plotted as a black diamond in figure~\ref{fig:ExEzRef425}($a$) and ($b$). We can see in figure~\ref{fig:ExEzRef425}($a$) that $\tau_z$ is close to the time when the decay of $E_x$ becomes faster, i.e. changes from one stage to another. This can be explained by the observation that rolls are present during the first stage of the decay, but have a negligible amplitude in the second stage (after $\tau_z$). The rolls generate streamwise perturbations in the form of streaks, which is called lift-up effect \citep{PJSchmid2001}. The decay of the streamwise component is sensitive to the presence of the other components. This effect has been discussed in particular in \cite{Rolland2015} who expresses the energy budget during the quench (equation~(9)) as the sum of a term linked to the interaction between streaks and rolls, and a term associated with the viscous dissipation of the streaks.

The magenta dashed line in figure~\ref{fig:ExEzRef425}($b$) represents an exponential fit of the function $E_z=E_0\exp(A_zt)$, where $A_z$ is the decay rate of $E_z$ and $E_0$ is the initial energy. This illustrates the energy $E_{z}$ decays exponentially under quenching. We initiated the fit $2$ data points (about $\Delta t\in[9,16]$) after $t=0$ to obtain a better fit as it reduces the sum of the squared residuals. This exponential fit covers approximately one decade of energy. The decay rate $A_z$ is linked to the decay time $\tau_z$ by the relation $A_z \approx \ln(0.05)/\tau_z$  (the decay of $E_z$ is not perfectly exponential, so the equality is only approximate).

The turbulent fractions $F_x$ and $F_z$ are plotted as a function of time in figure~\ref{fig:FtxFtzRef425}. The decay of $F_x$ also contained evidence for two different stages. We fit the decay of $F_z$ and the second decay stage of $F_x$ by a linear function $F_i = a_it+b$ ($i=x,z$). The best fit of the second slope $a_{x}$ of $F_x$ was obtained from a point just after $\tau_z$ to the time when the minimal slope was found with a minimum $12$ data points fitted. As a result of the limited data range of the second decay stage of $F_z$, both linear and exponential relationships provide acceptable fits. For consistency and simplicity in the rest of the paper, we use a  linear fit. The fits for the decay rate $A_z$ and decay slope $a_i$ are performed on 5 realisations, separately. The average and standard deviation of their values are presented and discussed in section~\ref{Variation of characteristic time with $Re$}.

The decay slope $a_{x}$ is greater than $a_z$, which means the rolls decay faster than the turbulent and laminar streaks. The change of slopes with $\Rey_f$ will be discussed in section ~\ref{Variation of characteristic time with $Re$}.  One hypothesis of linear decay of turbulent fraction is the formation of laminar holes and the linear increase of laminar region \citep{Rolland2015}. As \cite{Rolland2015} notes, numerical simulations of quenches in PCF show that the decay is exponential for the kinetic energy and linear for the turbulent fraction $F_t$ in the range $(\Rey_g,\Rey_t)$. We found linear decay is also valid for $\Rey_f<\Rey_g$. In addition, the two decay regimes of the streaks were revealed.

The equivalent plots to Figs.~\ref{fig:SnapRef425}, \ref{fig:ExEzRef425}, and \ref{fig:FtxFtzRef425} are shown in figures \ref{fig:SnapRef600}, \ref{fig:ExEzRef600} and \ref{fig:FtxFtzRef600} for the case of $\Rey_f=600$. It can be seen in figure~\ref{fig:ExEzRef600} that $E_{x}$ and $E_{z}$ suddenly decrease which indicates that the flow has changed to a less turbulent state (smaller $F_x$).  
As the decay is rapid and the number and range of data points is limited, an exponential fit does not provide a good fit to the data.
On the other hand, the decay times $\tau_x$ and $\tau_z$ are always well-defined and can be used to quantify the decay over a wide range of Reynolds numbers. Therefore, we used them in section~\ref{Variation of characteristic time with $Re$} to study the influence of $\Rey_f$ on the decay process.

After some time, the turbulent patches were advected away  from the measurement window towards reservoir $2$ (see figure~\ref{fig:SnapRef600}($c$) and ($g$)). We also observed that the streaks can re-enter the observation area from reservoir $1$. This can be observed in the snapshots of figure~\ref{fig:SnapRef600}($d$) and ($h$) and help explain the local maximum in energy at $t\approx 2430$ in figure~\ref{fig:ExEzRef600}. We will discuss these effects in detail in section~\ref{Advection of turbulent spots}, where we show that the first stage of the decay discussed here is not affected by this noise.

The snapshots of $u_x$ and $u_z$ at $\Rey_f=600$ shown in figure~\ref{fig:SnapRef600} illustrate a different decay scenario from the $\Rey_f=425$ case. The temporal evolution of $F_x$ and $F_z$ for $\Rey_f=600$ is shown in figure~\ref{fig:FtxFtzRef600} plotted on a $lin-lin$ scale. The turbulent fraction evolution is close to the energy evolution at $\Rey_f=600$. After quenching, as in figure~\ref{fig:SnapRef600}($b$), it can be seen that the streaks in the lower half part of the measurement window become elongated and straighten, in contrast to the turbulent streaks in the middle. At the same time, the rolls in the lower part become weak and the flow is approximately laminar. This means the long straight streaks cannot reinject energy into rolls. This observation is consistent with the mechanisms of reinjection of energy into the rolls driven by nonlinear interactions between wavy streaks \citep{Waleffe1997SSP}. The patch of streaks and rolls are subsequently advected by the moving wall towards reservoir $2$ as in figure~\ref{fig:SnapRef600}($c$). In figure~\ref{fig:SnapRef600}($d$), the streaks and rolls re-enter the measured field from the left side (i.e. from reservoir $1$).

\begin{figure}
	\centerline{\includegraphics[width=\textwidth]{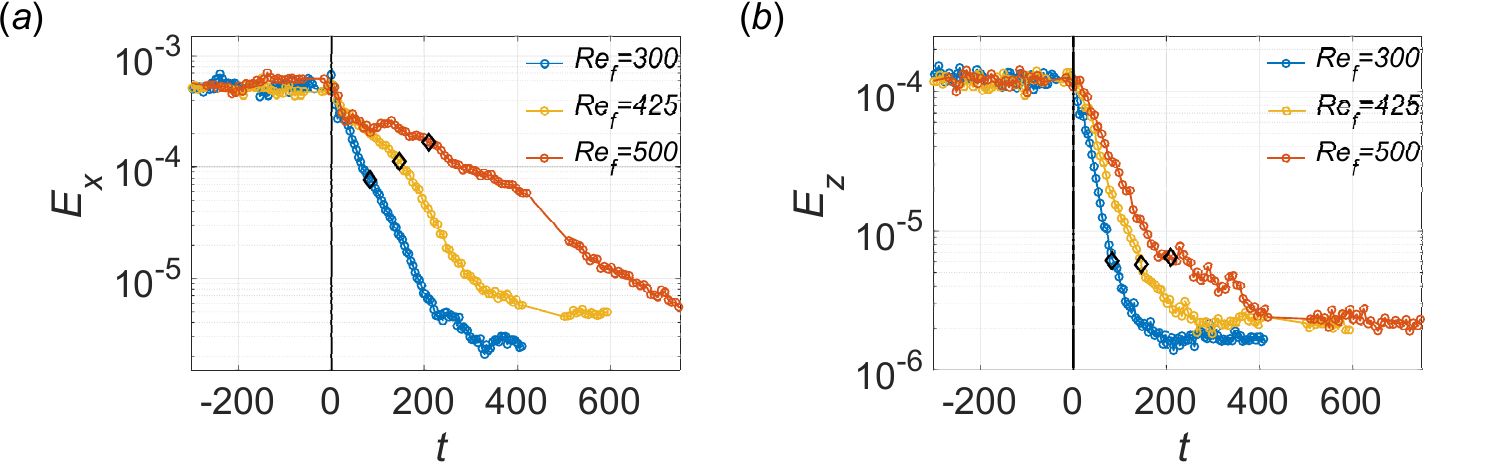}}
	\caption{Temporal evolution of $E_x$ in ($a$) and of $E_z$ in ($b$) for $\Rey_f=300$, $425$ and $500$; black diamonds: $\tau_z$ for each $\Rey_f$; noise intensity: high ($\sigma=4.6\times10^{8}$, defined in Section ~\ref{External noise in the permanent regime}).}
	\label{fig:ExEzRef300425500}
\end{figure}

\begin{figure}
	\centerline{\includegraphics[width=\textwidth]{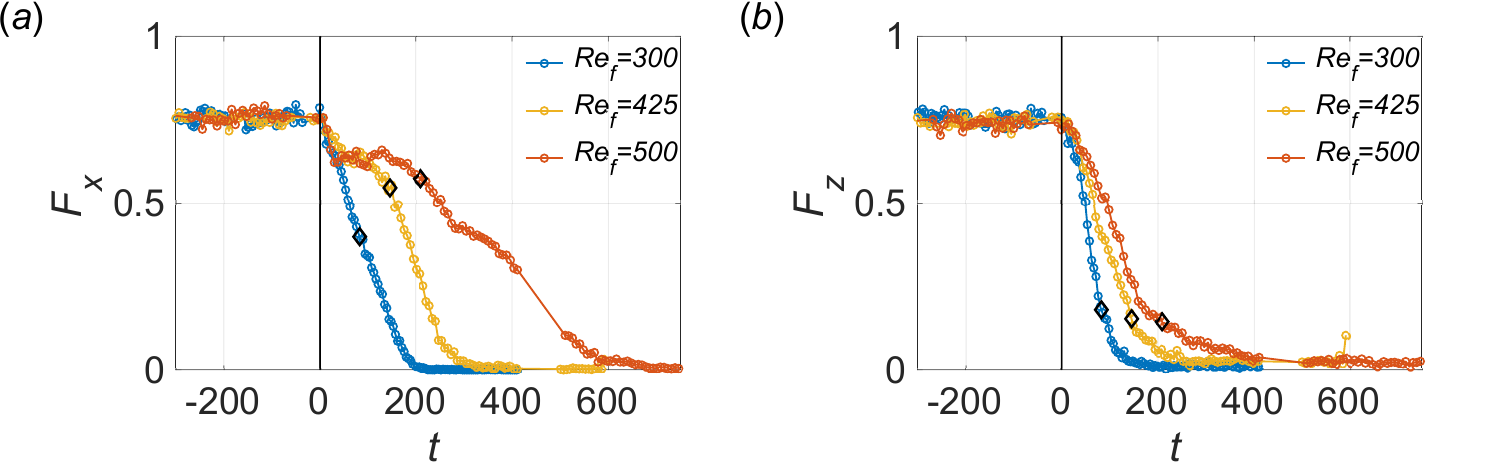}}
	\caption{Temporal evolution of $F_x$ in ($a$) and of $F_z$ in ($b$) for $\Rey_f=300$, $425$ and $500$; black diamonds: $\tau_z$ for each $\Rey_f$; noise intensity: high ($\sigma=4.6\times10^{8}$, defined in Section ~\ref{External noise in the permanent regime}).}
	\label{fig:FxFzRef300425500}
\end{figure}

In order to investigate the influence of lift-up effect for different $\Rey_f$, the energy evolution of $E_x$ and $E_z$ and the turbulent fraction evolution of $F_x$ and $F_z$ for three different final Reynolds numbers: $\Rey_f=300$, $\Rey_f=425$ and $\Rey_f=500$ are compared in figure~\ref{fig:ExEzRef300425500} and~\ref{fig:FxFzRef300425500}. The change of decay regime is not observed in the evolution of $E_x$ and $F_x$ for $\Rey_f=300$. This implies the lift-up effect at this small Reynolds number is not pronounced. As the $\Rey_f$ increase to $425$, we observe the existence of two decay stages. With the further increase of $\Rey_f$ to $500$, we observed that the energy $E_x$ and turbulent fraction $F_x$ first drop to a lower plateau after the quench and the plateau is maintained to until approximately $\tau_z$ when $E_z$ and $F_z$ decay to very low levels. This transient plateau is a result of the lift-up mechanism and the roll is a key ingredient. When the roll is no longer active, the plateau is not sustained.

In summary, we have uncovered important details of the decay process at $\Rey_f=425$ and $\Rey_f = 600$, respectively. The decay of turbulence is direct throughout the flow field at $\Rey_f=425$, in contrast to a partial decay or a formation spatially distinct laminar holes at $\Rey_f=600$. We made the observation that the decay rates and decay slopes are different by comparing the decays of the streamwise energy $E_x$ and the turbulent fraction $F_x$ with the spanwise energy $E_z$ and the turbulent fraction $F_z$, respectively. The decay of the streamwise component revealed two different decay stages depending on the presence of the roll component which is an important ingredient of the lift-up effect.

\section{Noise}
\label{Noise}

\subsection{External noise in the permanent regime}
\label{External noise in the permanent regime}

Noise is inevitably present in the experiment since there is a rotating cylinder driving a moving belt through a reservoir. Here we have varied the noise level using grids at the entrance to reservoir~$1$.  The efficiency of the grids depends on the mechanical mount supporting them and this was found to have a significant effect on the level of noise. In this section, we discuss measurements to illustrate that the level of noise could be controlled and quantified. The quantification is indirect, since the velocity field is the response of the flow field to the external noise. As mentioned in section~\ref{Decay process} and in the work of \cite{Kreilos2016} for boundary layer flow, the turbulent state is only observed when $z$ component is significant. In the following of this section, the noise levels are quantified using the roll component.

The time averaged spanwise energy of permanent state is: 
\begin{equation}
<E_{z}> = \frac{1}{t_b-t_a}\int_{t_b}^{t_a} E_{z} dt
\label{Eq:meanAzDefinition}
\end{equation}
where $t_a$ is the time when the transient decay ends after quenching, $t_b$ is the end of the measurement. We used $t_a=1500 > t_{adv}$, where $t_{adv}$ is the advection time during which the streaks travel from the entrance to the channel past the measured station (see section~\ref{Advection of turbulent spots}). In order to ensure the average started after the transient decay. $<E_{z}>$ is approximately a constant when the time span verifies $t_b-t_a>5 \times 10^3$.
The variance of the permanent state is defined by:
\begin{equation}
\chi_{z} = \sqrt{\frac{1}{t_b-t_a}\int_{t_a}^{t_b} ({E_{z}}^2-<E_{z}>^2)dt}
\label{Eq:VarianceAz}
\end{equation}

\begin{figure}
  \centerline{\includegraphics[width=0.7\textwidth]{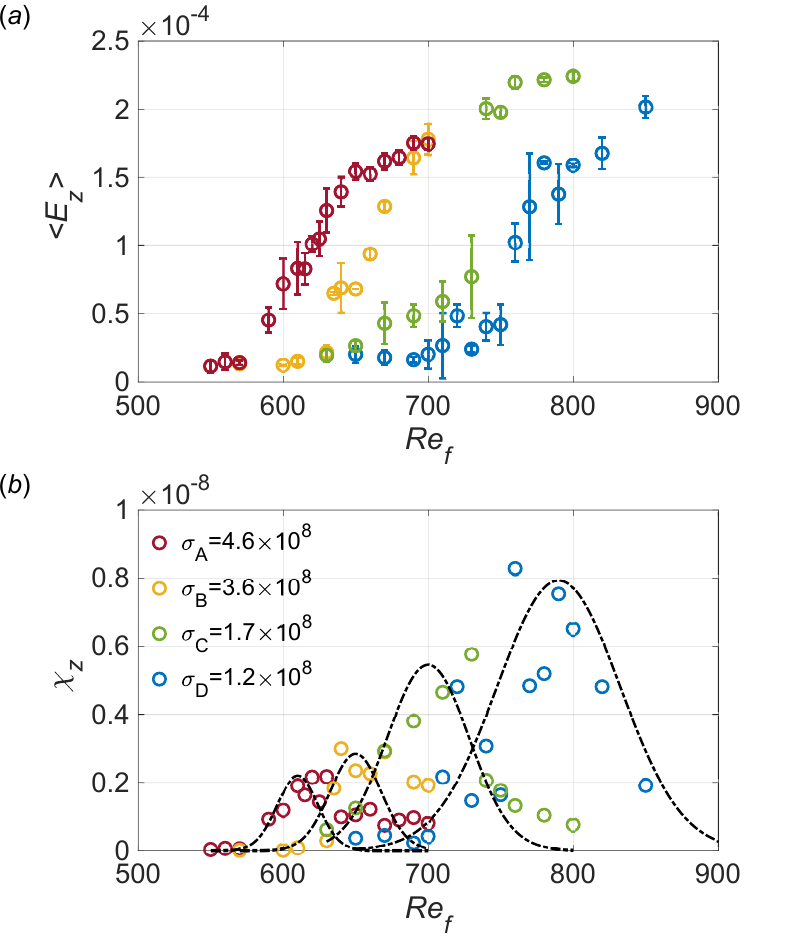}}
  \caption{(a) Time and space averaged amplitude $<E_{z}>$ of the final state as a function of $\Rey_f$ for different noise levels; error bar: standard deviation of 5 realisations for red and green data, 2 realizations for yellow and blue data. (b) Variance $\chi_z$ of the final state for different noise levels; black dot-dashed lines: guide for the eyes.}
\label{fig:VarianceAmplitudeSummary}
\end{figure}

\noindent A plot of $<E_{z}>$ as a function of $\Rey_f$ for the four different noise levels is given in figure~\ref{fig:VarianceAmplitudeSummary}($a$). The red points correspond to the experiments without grids, i.e. for which the noise is the greatest. The three other colors correspond to three different positions of the grid. The time averaged $<E_z>$ is linked to both the dynamics and the noise level and provides a measure of the response of the flow to the noise.

The different curves have a similar shape but are shifted along the $\Rey_f$ axis. The laminar state is linearly stable in this system and the noise is amplified through transient growth \citep{klotz_wesfreid_2017jfm}. It is thus expected that the greater the noise level, the higher the energy of the flow at a given $\Rey_f$.
From the figure, we rank the datasets high to low: red, yellow, green, blue, respectively. We observed that the flow remains laminar at $\Rey_f=680$ for the blue curve, i.e. the lowest noise level. 

We also characterised the noise using the variance. The idea of using the variance is inspired by the use of susceptibility (see for instance \cite{Garcia-Ojalvo}), where the external field would be replaced here by the noise. It is also inspired by \cite{Agez2013BifurcationNoise} and \cite{Rolland_2018}, who uses response functions to characterise  bifurcations in PCF. The noise is intrinsic in the case of \cite{Rolland_2018}, induced by the turbulence, whereas here we characterise it as an external disturbance.

The variance $\chi_{z}$ as a function of $\Rey_f$ for the different noise levels is presented in figure~\ref{fig:VarianceAmplitudeSummary}($b$). We observe that the maximum $\chi_{z}$ increases as the noise level decreases. Therefore, we define the inverse $\sigma$ of the maximum $\chi_{z}$ as a proxy of the noise intensity:
\begin{equation}
\sigma = \frac{1}{\max(\chi_{z})}
\label{Eq:NoiseIntensity}
\end{equation}

\begin{table}
	\begin{center}
		\def~{\hphantom{0}}
		\begin{tabular}{lccc}
			Noise level  &   Noise intensity $\sigma$ & Marker colour \\[3pt]
			High   & $\sigma_A = 4.6\times10^{8}$ & red\\
			Medium   & $\sigma_B =3.6\times10^{8}$ & yellow\\
			Low  & $\sigma_C = 1.7\times10^{8}$ & green\\
			Low   & $\sigma_D=1.2\times10^{8}$ & blue\\

		\end{tabular}
		\caption{List of noise levels and intensities; marker colour refers to the colour of the data points in figure 10, 14 and 15.}
		\label{tab:noiseintensity}
	\end{center}
\end{table}

\noindent We obtain the four noise intensities and corresponding noise levels which are listed in table \ref{tab:noiseintensity}.  We use the notation high ($\sigma_A$), medium ($\sigma_B$) and low noise ($\sigma_C$ and $\sigma_D$) levels throughout the paper to  indicate the various noise intensities defined here. The dominant frequency of the noise is close to the frequency of the belt motion loop.

The apparent threshold of CPF is shifted to higher $\Rey_f$ through reducing the noise level. This observation is similar to the work by \cite{Agez2013BifurcationNoise}. They use an amplitude equation model with additive noise to study the influence of the noise level on a sub-critical bifurcation. They report that the increase of the intensity of the additive noise shifts the threshold to lower values, similar to the imperfection sensitivity in shell buckling.

\subsection{Advection of turbulent spots}
\label{Advection of turbulent spots}

\begin{figure}
	\centerline{\includegraphics[width=\textwidth]{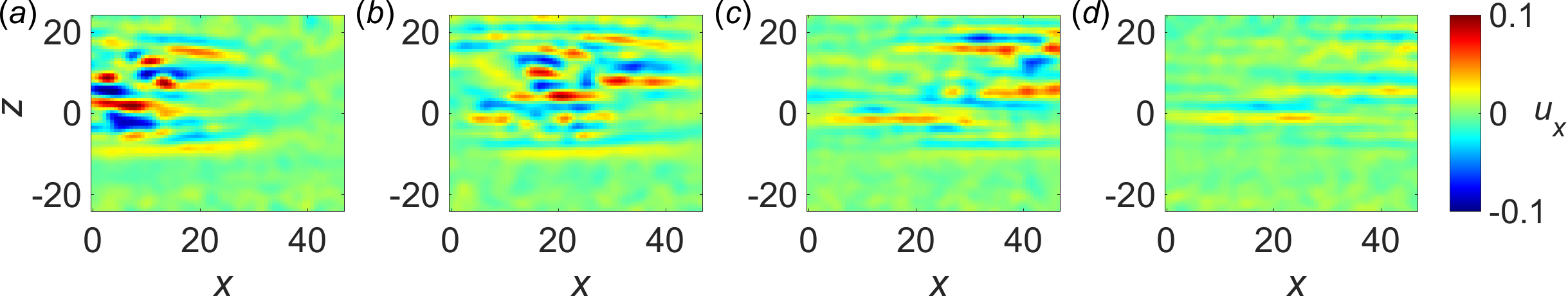}}
	\caption{Snapshots of $u_x$ at $(a)$ $t=2000$; $(b)$ $t=2100$; $(c)$ $t=2200$ and $(d)$ $t=2300$ for $\Rey_f =610$ ; noise level: medium ($\sigma_B$).}
	\label{fig:Snapshot_spot}
\end{figure}

\begin{figure}
  \centerline{\includegraphics[width=12cm]{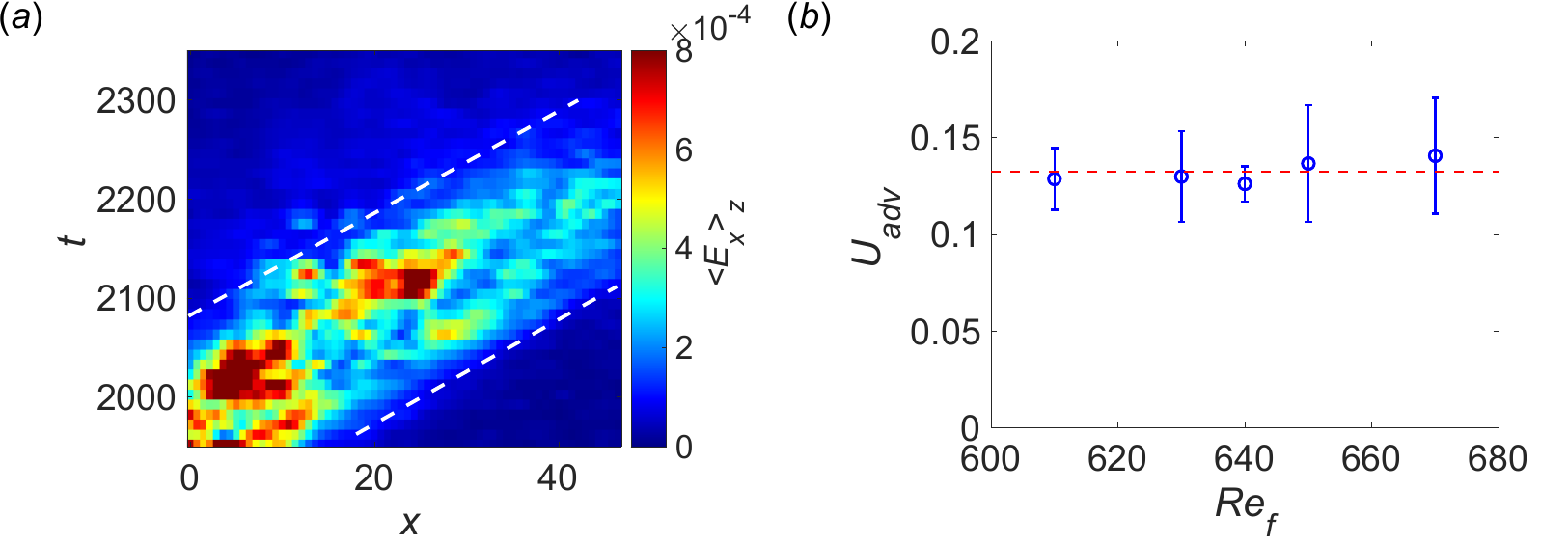}}
  \caption{(a) Spatio-temporal diagram of streamwise amplitude $<E_{x}>$ averaged over $z$ direction for $\Rey_f=610$ in the time range $t\in [1900, 2350]$ during which a patch of streaks is advected in the measured field, white dashed line: the separation between laminar flow and the front of the streaks. (The diagram corresponds to the figure~\ref{fig:Snapshot_spot}); (b) Estimated advection velocity of turbulent streaks as a function of $\Rey_f$, blue circle: estimating of $U_{adv}$ from the spatio-temporal diagram of $<E_{x}>_z$ (the slope of the white dashed line); red dashed line: mean $<U_{adv}>=0.13$ of the blue circles; error bar: standard deviation of 5 estimations. }
\label{fig:SpatialTemporalDiag_Uadv}
\end{figure}

Turbulent spots are observed in the permanent regime for Reynolds numbers close to the global stability threshold. An example of such a spot can be seen at $\Rey_f=600$ in figure~\ref{fig:SnapRef600}($d$, $h$). Further, its accompanying signature in the integral measurements, e.g. the clear bump around  $t=2430$ in the turbulent fraction is shown in figure~\ref{fig:FtxFtzRef600}. We examine now the advection of spots, which will be helpful to interpret the results in section~\ref{Variation of characteristic time with $Re$} concerning the variation of characteristic times with the Reynolds number which are independent of the external noise level. 

The advection of turbulent spots at $\Rey_f=610$ is shown in the series of snapshots in figure~\ref{fig:Snapshot_spot}. We observe that the spots are advected from left to right in figure~\ref{fig:Snapshot_spot} ($a-c$) and decay from figure~\ref{fig:Snapshot_spot}($c$) to ($d$). The observation that the streaks travel suggests that it is induced by the small mean velocity in the channel, the invasion of the turbulence and the asymmetric Couette-Poiseuille flow profile. The corresponding spatio-temporal diagram of the streamwise energy averaged over $z$ direction $<E_{x}>_z$ is plotted in figure~\ref{fig:SpatialTemporalDiag_Uadv}($a$) in order to study the evolution of a spot. We estimate the advection velocity of streaks $U_{adv}$ by the slope of the white dashed lines in figure~\ref{fig:SpatialTemporalDiag_Uadv}($a$) which separates the laminar flow and the streaks. These two lines are almost parallel, which suggests that the turbulent spots are advected and decay. We can observe from the energy evolution between the white dashed lines that the decay the spots are mainly due to the decrease of the energy without apparent reduction of the turbulent area. This corresponds to the fading trajectory and the minimal spot reported by \cite{Monchaux2020jfm}. $U_{adv}$ as a function of $\Rey_f$ is shown in figure~\ref{fig:SpatialTemporalDiag_Uadv}($b$). It is clear that the mean value of $U_{adv}$ is approximately constant with $<U_{adv}>=0.13$. (red dashed line in figure~\ref{fig:SpatialTemporalDiag_Uadv}($b$)).

The observation that the main source of the external noise is the rotating cylinder in reservoir $1$ (see figure~\ref{fig:setup}). The noise generates turbulent streaks and rolls that are advected from the entrance of reservoir $1$ towards reservoir $2$. We estimated the time for the streaks to be advected from the entrance of reservoir $1$ to the center of the measurement window as $t_{adv}=L/<U_{adv}>=1360$. As a result, if the decay time is longer than $t_{adv}$, the streaks and rolls will be advected away from the measured field. The measurement of the decay time $\tau$ is thus limited by $t_{adv}$. Our measurements of the decay time is smaller than this typical time $t_{adv}$. This implies that the measurements are not affected by the external noise.
However, the noise plays a major role in the permanent regime.

\subsection{Intrinsic noise}

\begin{figure}
  \centerline{\includegraphics[width=14cm]{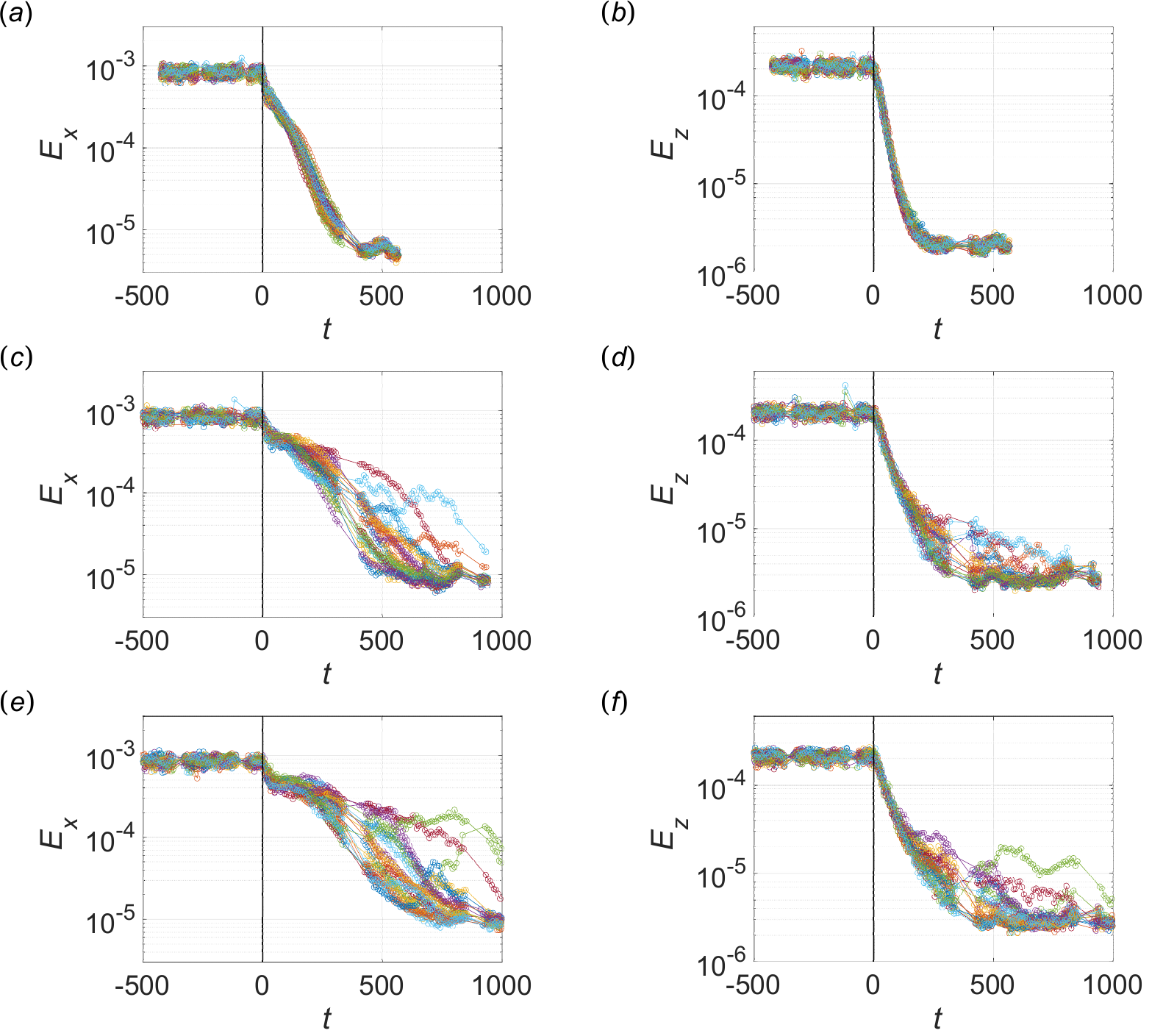}}
  \caption{Temporal energy evolution $E_x$ on the left and $E_z$ on the right for 20 quench experiments realisations at $\Rey_f=375$ in ($a$) and ($b$), $\Rey_f=510$ in $(c)$ and $(d)$, $\Rey_f=525$ in ($e$) and ($f$); noise level: low ($\sigma_C$).}
\label{fig:20quenchexperiments}
\end{figure}

As discussed in section~\ref{Advection of turbulent spots}, the external noise has no detectable influence on the transient decays. However, since the flow is turbulent, we observe some variability between each realization of the quench procedure. We find that five realizations of each quench are sufficient to enable a meaningful average decay time.

The energy evolution of the streamwise component (streaks) $E_{x}$ is shown in figure~\ref{fig:20quenchexperiments} in the left column and of spanwise component (rolls) $E_{z}$ in the right column for $20$ repeated realisations for $\Rey_f=375$, $510$ and $525$ with low noise level ($\sigma_C$), respectively. The energy evolution of $E_{z}$ is repeatable for $\Rey_f=375$. For instance in figure~\ref{fig:20quenchexperiments}($a$) and ($b$), the relative variation (ratio of the standard deviation to the mean value) of $\tau_x$ and $\tau_z$ are $9\%$ and $8\%$ for $\Rey_f=375$, respectively. When $\Rey_f>500$, the energy evolution of $E_{x}$ begins to spread and is different for each realization. The relative variation of $\tau_x$ and $\tau_z$ are $31\%$ and $26\%$ for $\Rey_f=525$ in figure~\ref{fig:20quenchexperiments}($e$) and ($f$), respectively. The spread of the realisations increases with $\Rey_f$. However, even at the highest $\Rey_f$, the spread is small. Note that this decay time is different from the lifetimes obtained from the probability distribution of the relaminarization times, which is wide close to $\Rey_g$ \citep{Celso1986PRL,BottinChate1998}.

\section{Variation of characteristic times with $\Rey_f$}
\label{Variation of characteristic time with $Re$}

In this section, we discuss the variation with the final Reynolds number of: ($1$) the decay times obtained from the energy curves, and ($2$) the decay rates $A_z$ obtained from exponential fits of $E_z$ (see figure~\ref{fig:ExEzRef425}($b$)) and ($3$) the decay slopes $a_x$ of the second decay stage of $F_x$ and $a_z$ of $F_z$ (see figure~\ref{fig:FtxFtzRef425}($a$) and ($b$)).

\begin{figure}
	\centerline{\includegraphics[width=\textwidth]{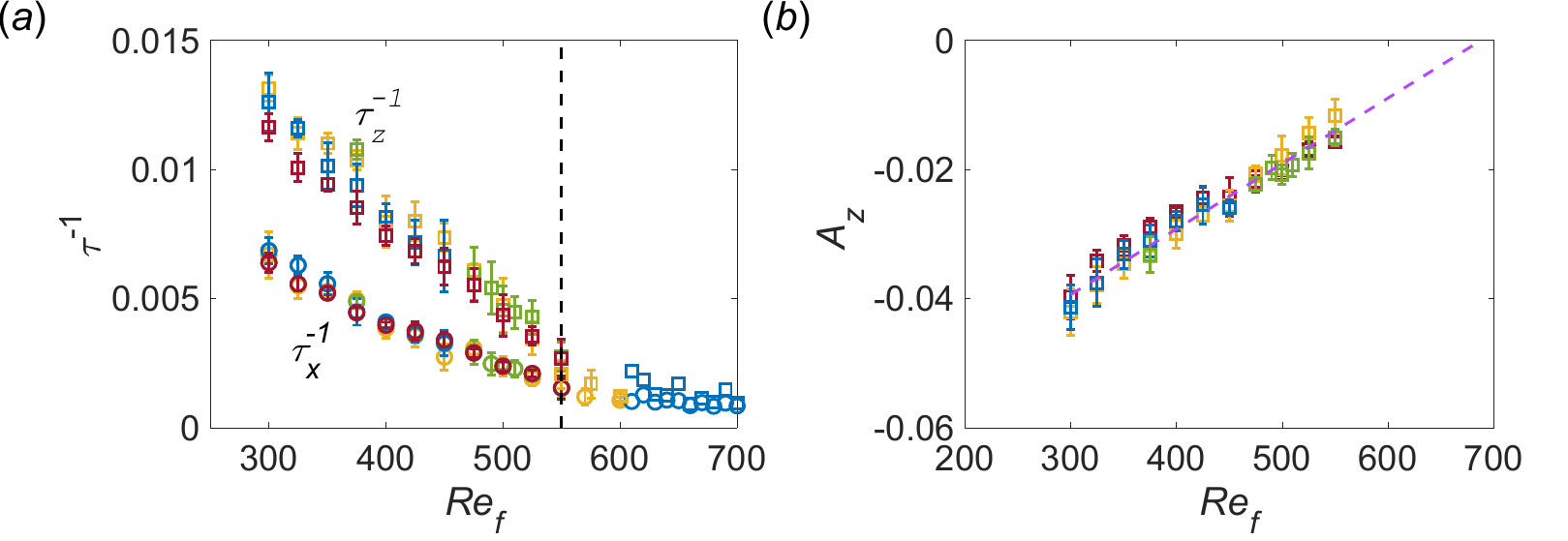}}
	\caption{($a$) $\tau^{-1}$ as a function of $\Rey_f$ for four different noise levels, red: high($\sigma_A$); yellow: medium($\sigma_B$); green: low($\sigma_C$); blue: low($\sigma_D$) extracted from the energy evolution of $E_{z}$ (circle) and $E_{z}$ (square); error bar: standard deviation of several realisations. $\tau$ is defined at the time when energy decays to $5\%$ of its initial level (see section~\ref{Decay process}). ($b$) Decay rate $A_z$ of the spanwise energy evolution $E_z$ as a function of $\Rey_f$; color represents noise levels: red: high($\sigma_A$); yellow:~medium($\sigma_B$); green:~low($\sigma_C$); blue:~low($\sigma_D$); purple dashed lines: linear fit of the mean decay slopes; error bar: standard deviation of $5$ realisations.}
	\label{fig:tauinversesummary}
\end{figure}

The inverse of the decay times $\tau_x^{-1}$ and $\tau_z^{-1}$ (defined in section~\ref{Decay process}) are plotted as a function of $\Rey_f$ for four different noise levels in figure~\ref{fig:tauinversesummary}($a$). The decay time $\tau_{x/z}$ is defined as the time when the energy of the streaks $E_x$ or rolls $E_z$ decrease to a threshold, which we set at $5\%$ of its initial value (see Section~\ref{Decay process}). For $\Rey_f\geq550$, we observed that the decay is not fully captured in the measurement and there are turbulent patches which travel out of the measurement field at the speed of the advection velocity. This advection leads to a saturation of $\tau$ at $t\gtrsim 1100\approx t_{adv}$ when $\Rey_f\gtrsim 600$. When the noise level is relatively high (ex. noise intensity $\sigma_A$ and $\sigma_B$), $\tau$ cannot be defined for these high $\Rey_f$ as the flow is never in a completely laminar state, so that only data corresponding to low noise are displayed in figure~\ref{fig:tauinversesummary}($a$) (blue points) for $\Rey_f\gtrsim 600$. In the rest of this section, we focus on the decay for $\Rey_f<550$. We observed that the inverse of the decay times $\tau_x^{-1}$ and $\tau_z^{-1}$ decrease when $\Rey_f$ increases. The decay time $\tau_x$ is always greater than $\tau_z$ and the ratio $\tau_x/\tau_z$ is between $1.7$ and $2.0$. This means $E_z$ decays faster than $E_x$ over the whole range of $\Rey_f$. The values of $\tau_x$ and $\tau_{z}$ are found to be independent of the different noise levels. 

We have also carried out a series of investigations into the choice of the threshold for the definition of $\tau$ (see section~\ref{Decay process}). The decay times $\tau_x$ and $\tau_z$ presented in figure~\ref{fig:tauinversesummary}($a$) are defined as the times when the energy decays to a threshold of $5\%$ of the initial energy. The threshold was varied between $5\%$ and $18\%$. At the largest threshold of $18\%$, $\tau_x$ is only valid over the first stage of the decay (see figure~\ref{fig:FtxFtzRef425}) of $F_x$, whereas $\tau_x$ includes both the first and a part of second decay stage of $F_x$ with a smaller threshold, e.g. $5\%$. The decay time $\tau_x$ is greater than $\tau_z$ irrespective of the threshold. This indicates a faster decay rate of spanwise energy applies during both the first decay stage and the whole decay. To conclude, the measurements of the decay times of $E_x$ and $E_z$ confirms the $E_x$ decays slower than the $E_z$ presented in section~\ref{Decay process} for all $\Rey_f<550$ regardless of the noise levels.

\begin{table}
	\begin{center}
		\def~{\hphantom{0}}
		\begin{tabular}{lccc}
			Noise level  &   $\Rey_{A_z}$ & $\Rey_{a_z}$\\[3pt]
			High   & $\Rey_{A_z}=711\pm26$ & $\Rey_{a_z}=659\pm14$  \\
			Medium   & $\Rey_{A_z}=668\pm17$ & $\Rey_{a_z}=640\pm9$  \\
			Low  & $\Rey_{A_z}=701\pm13$ & $\Rey_{a_z}=674\pm10$ \\
			All data & $\Rey_{A_z}=688\pm10$ & $\Rey_{a_z}=656\pm10$ \\
			
		\end{tabular}
		\caption{List of crossover Reynolds numbers $\Rey_{A_z}$ and $\Rey_{a_z}$ obtained from the linear extrapolation of decay rates and decay slopes as a function of $\Rey_f$ at different noise levels.}
		\label{tab:crossoverRe}
	\end{center}
\end{table}

The decay of $E_z$ after the quench can be well-fitted by an exponential function $E_z=E_0\exp(A_zt)$ (see Fig~\ref{fig:ExEzRef425} ($b$)). The decay rate $A_z$ is plotted as a function of $\Rey_f$ in figure~\ref{fig:tauinversesummary}(b) for all noise levels. This decay rate decreases as $\Rey_f$ is increased and scales linearly with the $\Rey_f$. The variation of $A_z$ with $\Rey_f$ is fitted with the function $A_z\propto(\Rey_f-\Rey_{A_z})$. The crossover Reynolds numbers $\Rey_{A_z}$ obtained from this linear extrapolation are listed in table~\ref{tab:crossoverRe} for different noise levels (uncertainty is obtained using the bootstrap method).  All these crossovers $\Rey_{A_z}$ are consistent with $\Rey_{A_z}=688\pm10$ obtained by fitting $A_z$ using data taken over the range of noise levels.

\begin{figure}
	\centerline{\includegraphics[width=0.55\textwidth]{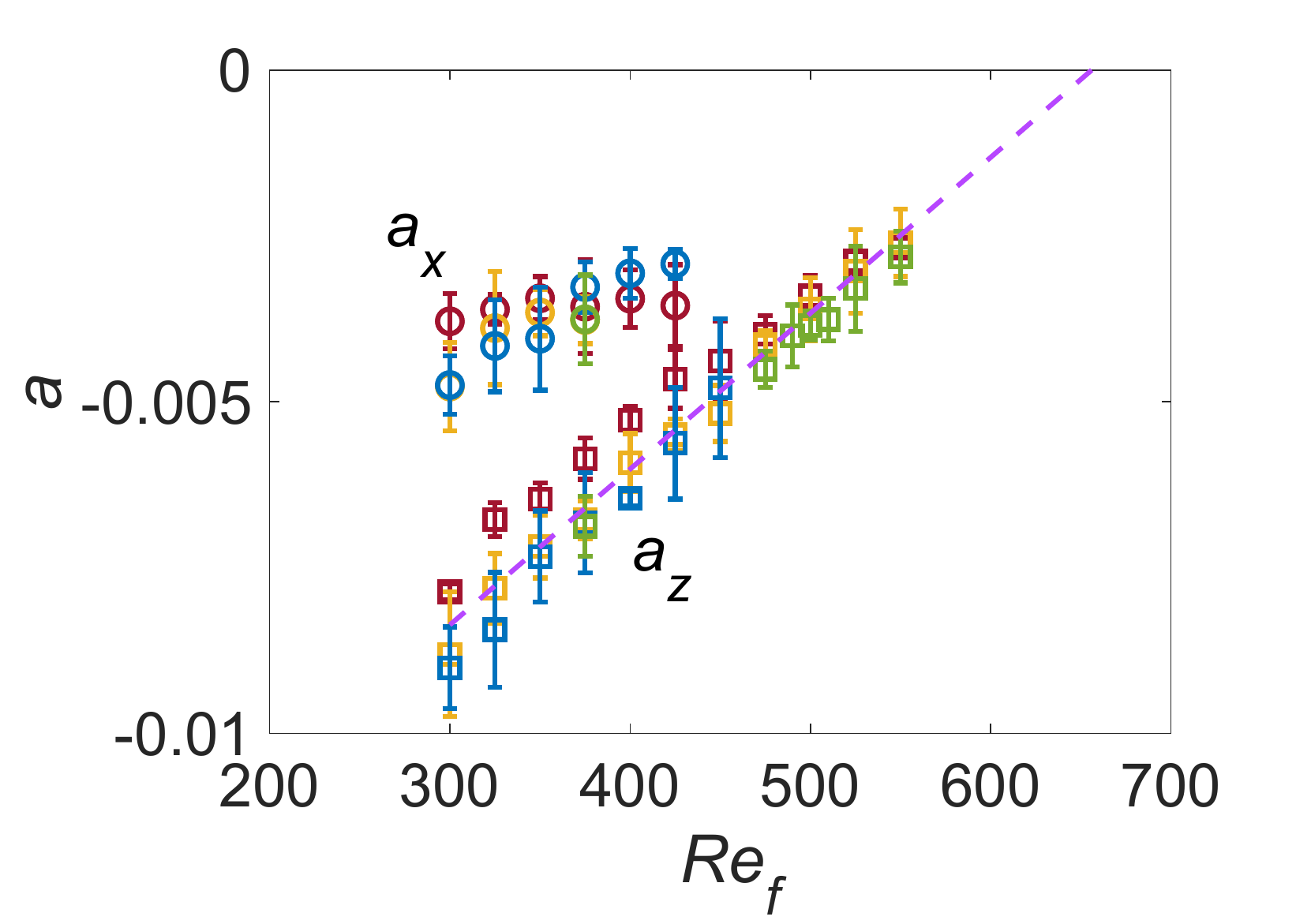}}
	\caption{Circles: slope $a_{x}$ of the decay of turbulent fraction $F_x$ after $\tau_z$ ($F_z$ has nearly decayed at $\tau_z$) as a function of $\Rey_f$; squares: slope $a_{z}$ of the decay of turbulent fraction $F_z$ as a function of $\Rey_f$; color represents noise levels: red: high($\sigma_A$); yellow:~medium($\sigma_B$); green:~low($\sigma_C$); blue: ~low($\sigma_D$); purple dashed lines: linear fit of the mean decay slopes; error bar: standard deviation of $5$ realisations.}
	\label{fig:az_linearfit}
\end{figure}

In contrast to exponential decay of $E_z$, the spanwise turbulent fraction evolution $F_z$ has a linear decay after the quench. Similarly, the second decay stage of $F_x$ is also linear for $\Rey_f\leq425$ (see figure~\ref{fig:FtxFtzRef425}($a$) in section~\ref{Decay process}). The linear decays are fitted by the function $F_{i} = a_{i}t+b$ ($i=x,z$). The decay slopes $a_x$ and $a_z$ plotted as a function of $\Rey_f$ are shown in figure~\ref{fig:az_linearfit}, respectively. The slope $a_x$ is always greater than $a_z$ in the range of $\Rey_f\in[300,425]$ irrespective of the noise level. This means there is a faster decay along the spanwise direction. This finding is consistent with the observation that the second decay stage of $F_x$ decays slower than the decay of $F_z$ as can be seen in figure~\ref{fig:az_linearfit}. The slope $a_z$ scales linearly with $\Rey_f$ and is fitted by the function $a_z\propto(\Rey_f-\Rey_{a_z})$. The crossovers $\Rey_{a_z}$ obtained from the linear fit with the data from different noise levels are listed in table~\ref{tab:crossoverRe}. The crossover $\Rey_{a_z}$ fitting $a_z$ for all the different noise levels is found at $\Rey_{a_z}=656\pm10$ which is consistent with the value obtained from the fit of $a_z(\Rey_f)$ of each noise level.

The crossover Reynolds numbers obtained from the linear fit of $A_z(\Rey_f)$ and $a_z(\Rey_f)$ are close to the self-sustained threshold. This threshold is at approximately $\Rey\approx670$ \citep{Klotz2017PRF}. The linear scalings are obtained far from the threshold, in the range $\Rey_f\in[300,525]$.
This is different from many previous studies, which focused on the behaviour of characteristic times very close to the threshold (see for instance \cite{BottinChate1998, LiangShi2013PRL_PCF}). 
An example of the investigation of $\Rey_f$ far from $\Rey_g$ is given in \cite{Schneider2010PRE_PCF}. They define a characteristic time from the lifetime distribution, using direct numerical simulations in the PCF geometry. The points in the range $\Rey_f\in[250,280]$, which is far from the $\Rey_g\approx 325$, are well-fitted by a law $\tau^{-1} \propto(\Rey_g-\Rey_f)$ (figure~3 of \cite{Schneider2010PRE_PCF}. The linear scaling of characteristic times with $\Rey_f$ far from $\Rey_g$ provides an  estimate of the value of the global stability threshold. In our case, we have also observed linear scalings of decay rates and slopes with $\Rey_f$. At present, we do not have a theoretical explanation for the linear scalings.

\section{Conclusion}

We have investigated the decay of turbulence in Couette-Poiseuille flow using quench experiments. The final Reynolds number was varied over a wide range, and included all values where rapid decays are obtained. Thus we have extended the range of previous investigations which were primarily focused on values close to the critical point. We have provided experimental evidence for different decay rates for streaks and rolls during decay, which is consistent with previous theoretical work \citep{Waleffe1997SSP, Rolland2018PRE}. From the temporal evolution of the energy of streaks and rolls, we have shown that the rolls injects energy into the streaks through the lift-up mechanism during the first decay stage.

As in all plane channel experiments with moving walls, noise is generated in the reservoirs at the ends of the channel can induce turbulence in the flow. We have investigated the effects of this using multi-layer grids at the entrance (see figure~\ref{fig:setup}) to reduce the noise. We find that the characteristic decay time and decay rates are independent of the noise levels. However, the permanent regime after the transient decay is sensitive to the noise intensities which are quantified by the susceptibility of the spanwise roll energy. The mean amplitude of the energy at a given $\Rey_f$ increases with the noise which is a result of the advection of turbulent spots into the measurement field.

We have characterized the relaminarisation after a quench of an initially turbulent flow using the decay times, decay rates of the roll energy $E_z$ and the decay slopes of the turbulent fraction of the roll component $F_z$. The comparison of decay times provides evidence that the rolls always decay faster than the streaks by a factor around 2. The different decay rates are an essential feature in Walefffe's self-sustained model which was first investigated in a minimal flow unit with periodic boundary conditions \citep{Waleffe1997SSP}. Our experimental results support these ideas.

The decay of spanwise energy $E_z$ displays an exponential tendency, reminiscent of a viscous damping. This feature has previously been highlighted in numerical quench studies in PCF by \cite{Rolland2015}. The spanwise turbulent fraction $F_z$ displays a linear decay trend which has also been observed in numerical simulations of quench. The decay rates and decay slopes also contain a linear dependence on the $\Rey_f$, which is  independent of noise level. The theoretical explanation of this linear trend is not fully understood. The extrapolated value of $\Rey$ for which the decay rates and slopes vanish are close to the previously reported value for $\Rey_g\approx670$ in CPF \citep{Klotz2017PRF}.

The determination of the critical point $\Rey_g$ in Couette-Poiseuille flow using the crossing of the lifetime and splitting time has not been reported previously to our knowledge. However, it has been determined in PCF \citep{LiangShi2013PRL_PCF}, and in PPF \citep{Sebastien2020PRF}. It is thus very difficult to determine how close the extrapolated value obtained in our experiments are from $\Rey_g$. It is however noticeable that using a fit at Reynolds numbers much smaller than the critical point we can obtain a reasoned approximation.

The self-sustained process in channel flow is characterized by the presence of wavy streamwise streaks which lead to the breakdown of streaks and re-injection of energy into rolls due to non-linear effects (Waleffe 1997). In our measurements of the velocity field in section~\ref{Decay process}, we observed a fast extinction of the undulation of the streaks and the decay of the rolls followed by a slower decay of the straighten streaks. Our next step will be to measure the temporal evolution of the rolls and of the waviness of streaks simultaneously. This is the subject of an ongoing investigation and it will allow us to investigate the interplay between the three components of the self-sustained process.

\section*{Supplementary movies} Supplementary movies are available at https://doi.org/10.1017/jfm.2021.89.

\section*{Acknowledgements}

We gratefully acknowledge Joran Rolland, Yohann Duguet, Romain Monchaux, S\'ebastien Gom\'e, Laurette Tuckerman, Dwight Barkley, Olivier Dauchot and Sabine Bottin for fruitful discussions. We thanks Xavier Benoit-Gonin, Amaury Fourgeaud, Thierry Darnige, Olivier Brouard and Justine Laurent for technical help. 

\section*{Funding}

This work has benefited from the ANR TransFlow, and by starting grants obtained by B.S. from CNRS (INSIS) and ESPCI. T.M. was supported by a Joliot visiting professorship grant from ESPCI.

\section*{Declaration of interests}

The authors report no conflict of interest.

%
\bibliographystyle{jfm}

\end{document}